\documentclass[aps,prl,twocolumn,superscriptaddress,showpacs]{revtex4}

\usepackage{amsmath}
\usepackage{graphicx}

\newcommand{\be}{\begin{equation}}
\newcommand{\ee}{\end{equation}}
\newcommand{\mb}[1]{\mathbf{#1}}
\newcommand{\nn}{\nonumber}
\newcommand{\Journal}[4]{#1 \textbf{#2}, #3 (#4)}

\newcommand{\PRev}{Phys. Rev.}
\newcommand{\ho}{HoMnO$_3$}

\begin{document}

\title{Ferroelectricity in the Magnetic E-Phase of Orthorhombic Perovskites}

\author{Ivan A. Sergienko}
\affiliation{Materials Science and Technology Division, Oak Ridge
National Laboratory, Oak Ridge, TN 37831, USA}
\affiliation{Department of Physics \& Astronomy, The University of Tennessee,
Knoxville, TN 37996, USA}

\author{Cengiz \c{S}en}
\affiliation{National High Magnetic Field Laboratory and Department of Physics,
Florida State University, Tallahassee, FL 32310, USA}

\author{Elbio Dagotto}
\affiliation{Materials Science and Technology Division, Oak Ridge
National Laboratory, Oak Ridge, TN 37831, USA}
\affiliation{Department of Physics \& Astronomy, The University of Tennessee,
Knoxville, TN 37996, USA}

\begin{abstract} 
We show that the symmetry of the spin zigzag chain E phase of the orthorhombic perovskite manganites and 
nickelates allows for the existence of a finite ferroelectric polarization. The proposed 
microscopic mechanism is independent of spin-orbit coupling. We predict that the polarization induced by the 
E-type magnetic order can potentially be enhanced by up to two orders of magnitude with respect to that in the spiral 
magnetic phases of TbMnO$_3$ and similar multiferroic compounds. 
\end{abstract}

\pacs{75.80.+q, 75.50.Ee,  77.80.-e, 75.47.Lx}

\maketitle

\emph{Introduction.}
The switching of the electric polarization $\mb P$ by a 
magnetic field of a few Tesla discovered in 
TbMnO$_3$~\cite{Kimura03a} and TbMn$_2$O$_5$~\cite{Hur04} has ignited enormous interest in a class of 
materials that can be termed \emph{improper magnetic ferroelectrics} (IMF's). While there is no intrinsic 
ferroelectric (FE) instability in the IMF's, $\mb P$ emerges due to its coupling to the primary 
\emph{magnetic} order parameter. Hence, the FE phase transition coincides with the 
corresponding magnetic transition, and $\mb P$ is very sensitive to magnetic field induced changes of 
the magnetic state~\cite{Kimura03a,Hur04,Goto04,Kimura05a,Lawes05,Kimura05b,Chapon06,Kimura05c,Yamasaki06}.
Symmetry imposes rather strict conditions on the possible magnetic order parameters -- the magnetic 
structure must have low enough symmetry in order for the system to form a polar axis. 
As a consequence, the IMF's often have complicated noncollinear structures,
including spiral and 
incommensurate~\cite{Blake05,Kimura05c,Lawes05,Chapon06,Kenzel05,Arima06,Yamasaki06,Taniguchi06}, 
while the IMF phases with collinear magnetism are rare~\cite{Chapon04,Aliouane06}. Noncollinear 
magnetic structures are stabilized due to either competing interactions (frustration) or 
anisotropies generated by spin-orbit coupling, which leads to reduced transition temperatures and 
weaker order parameters. The magnitude of $P$ ($\approx 0.1\,\mu$C/cm$^2$) is also affected by its 
weak coupling to magnetism. 
In turn, collinear IMF's may prove more promising for future applications as they 
are less prone to the obstacles mentioned above.

In our present study we turn our attention to the \emph{collinear} E-type magnetic phase that has been observed in 
perovskite manganites~\cite{Munoz01,Zhou06} and nickelates~\cite{GM94,Alonso99,FD01,Zhou05}. 
\emph{First}, we show that this is a previously overlooked example of an IMF. \emph{Second}, in contrast to the most extensively 
theoretically studied case of spiral magnetism~\cite{Katsura05,Sergienko06,Mostovoy06},
the mechanism responsible for ferroelectricity in the magnetically ordered phase does not rely on the
presence of anisotropic Dzyaloshinskii-Moriya interaction. In our case, $\mb P$ appears due to a gain in
the band energy of the $e_g$ electrons in the double-exchange model. 
We estimate that $P$ can be potentially larger than that in the spiral magnets by up to two orders of 
magnitude, thus paving a way for IMF's to reach the values of $P$ common for other multiferroic 
systems with better ferroelectric properties~\cite{Ederer06,Fiebig05}.

\begin{figure}
\vspace{0mm}
\includegraphics[clip,width=85mm]{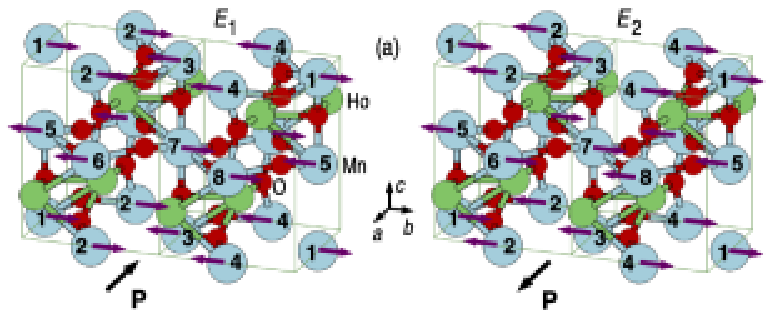}\\
\includegraphics[clip,width=75mm]{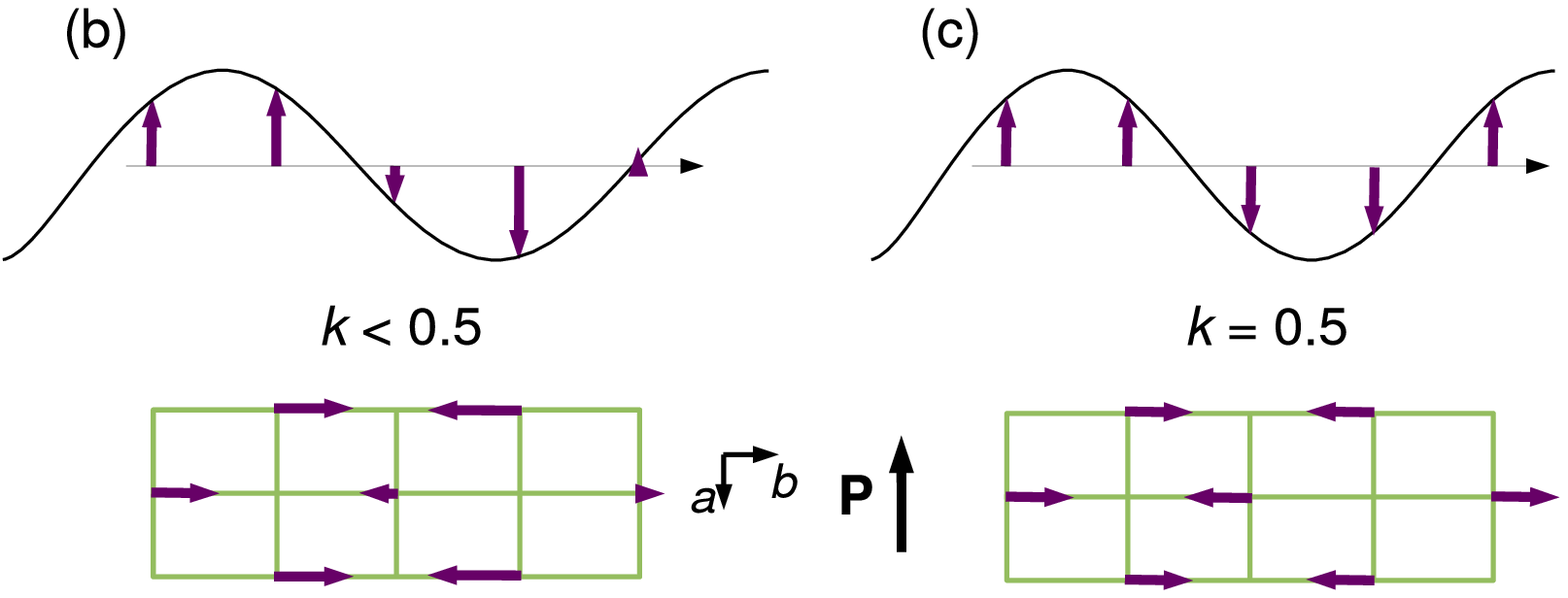}
\caption{\label{domains}(Color online) (a) Magnetic unit cells of the two E-phase domains in HoMnO$_3$ corresponding to 
the $(E_1,0,-P_a)$ and $(0,E_2,P_a)$ solutions of~(\ref{pot}). The arrows on the Mn atoms denote 
the directions  of their spins. The FE displacements are not shown, but the direction of $\mb P$ is 
indicated. (b) The simple sine-wave magnetic structure of \ho\ for $T_L < T < T_N= 42.2 - 47.5$~K~\cite{Munoz01}. 
(c) The E-phase magnetic structure below $T_L$.}
\end{figure}

In the perovskite manganite family $R$MnO$_3$ (space group $Pbnm$), the E-phase was first reported for $R$=Ho
as a result of magnetic structure refinement by neutron diffraction~\cite{Munoz01}. 
The magnetic unit cell of the E-phase, forming at low temperature below 
$T_L = 26 - 29.6$~K~\cite{Munoz01,Lorenz04}, is shown in Fig.~\ref{domains}(a). 
The Mn atoms with parallel spins form zigzag chains in the $ab$-plane, with the 
chain link equal to the nearest-neighbor Mn-Mn distance. The neighboring zigzag chains in the $b$-direction have 
antiparallel spins. Figures~\ref{domains}(b) and (c) show that the equal-spin E-phase structure can be obtained 
from the simple sine-wave structure by locking-in its modulation vector $k_b = \frac 1 2$ and fine-tuning its 
phase. The $ab$-planes are stacked antiferromagnetically ($+-+-$) along the $c$-direction.
Interestingly, Lorenz \emph{et al.}~\cite{Lorenz04} reported on a large magnetic field dependence of the dielectric 
constant below $T_L$, which may be an indication of the multiferroic state. 

\emph{Landau theory}.
Independently of the microscopic mechanism, the possibility of a FE state in magnets can be examined by 
considering the symmetry allowed terms in the Landau potential~\cite{Goshen70,Mostovoy06,Lawes05,Kenzel05,Harris06}. 
We define the symmetric coordinates corresponding to the E-phase as
\begin{eqnarray}
\label{Ecomb}
\mb E_1 &=& \mb S_1 + \mb S_2 - \mb S_3 - \mb S_4 - \mb S_5 - \mb S_6 + \mb S_7 + \mb S_8,\nn\\
\mb E_2 &=& \mb S_1 - \mb S_2 - \mb S_3 + \mb S_4 - \mb S_5 + \mb S_6 + \mb S_7 - \mb S_8,
\end{eqnarray}
where $\mb S_i$ is the spin of the $i$th Mn atom in the magnetic unit cell, as shown in Fig~\ref{domains}(a).
Since the Mn spins in \ho\ point along the $b$-axis~\cite{Munoz01}, below we consider only
the $b$-components of $\mb E_{1,2}$ denoted by $E_{1,2}$. However, the expression for the Landau 
potential derived below is valid for any component of $\mb E_{1,2}$.
$E_1$ and $E_2$ span an irreducible representation of the space group $Pbnm$ corresponding to 
$\mb k= (0 \frac 1 2 0)$. The properties of this representation are summarized in Table~\ref{Rep}. 
Taking into account that $\mb P$ transforms as a polar vector, we obtain the following form of 
the Landau potential corresponding to the E-phase,
\be\label{pot}
\begin{array}{l}
\displaystyle F = a (E_1^2+E_2^2) + b_1 (E_1^2+E_2^2)^2 + b_2 E_1^2E_2^2 \\
\displaystyle\quad + c (E_1^2-E_2^2) P_a + d (E_1^2-E_2^2) E_1 E_2 P_b + \frac 1 {2 \chi} \mb P^2,
\end{array}
\ee
where $\chi$ is the dielectric susceptibility of the paraelectric phase, and the other coefficients
are phenomenological parameters.
Minimizing $F$ with respect to $\mb P$, we obtain 
$P_a = - c \chi (E_1^2-E_2^2), \quad P_b = - d \chi (E_1^2-E_2^2)E_1E_2$,
and $P_c = 0$. Therefore, each of the four domains of the E-phase [$(\pm E_1,0)$ and $(0, \pm E_2)$] 
is IMF with the polarization along the $a$-axis and different signs of $P_a$ for $E_1$ and $E_2$.
We also notice that the $b$-axis component of $\mb P$ can be 
locally present within the domain walls where $E_1$ and $E_2$ coexist~\cite{MostNote}.

The E-type phase in the nickelates also consists of spin zigzag chains that 
have a different direction in the $ab$-plane and a different stacking along the 
$c$-axis ($++--$)~\cite{GM94,Alonso99,FD01,Zhou05}. 
The corresponding modulation vector is $\mb k = (\frac 1 2 0 \frac 1 2)$. The Landau theory analysis 
leads to results similar to the manganite case. For nickelates, the symmetric coordinates $E_1$ and $E_2$
(not given here) are written in terms of 16 Ni spins of the magnetic unit cell which is 4 times the 
crystallographic unit cell. It leads to a Landau potential similar to~(\ref{pot}) with $P_a$ replaced by $P_b$ 
in the fourth term and $d=0$. Thus, $\mb P$ in the E-phase of nickelates is parallel to the $b$-axis.

\begin{table}
\caption{\label{Rep}Matrices of the generators of space group $Pbnm$ in the irreducible representation
spanned by $E_1$, $E_2$. The space group elements are denoted $(r|hkl)$, where $r$ is the identity 
operation $1$, two-fold rotation $2_{a,c}$, inversion $I$, or time reversal $1'$ followed by the translation 
$\boldsymbol \tau = h\mb a + k\mb b + l \mb c$.}
\begin{ruledtabular}
  \begin{tabular}{cccc}
    & $(2_a|\frac 1 2 \frac 1 2 0)$ & $(2_c|0 0 \frac 1 2)$, $(I|000)$ & $(1|010)$, $(1'|000)$\\
    \hline
    $\begin{matrix}
      E_1 \\ E_2
    \end{matrix}$ & 
    $ 
    \begin{matrix}
      -1 & 0\\
      0 & 1
    \end{matrix}
    $ &
    $
    \begin{matrix}
      0 & 1\\
      1 & 0
    \end{matrix}
    $ &
    $ 
    \begin{matrix}
      -1 & 0\\
      0 & -1
    \end{matrix}
    $
  \end{tabular}
\end{ruledtabular}
\end{table}

\emph{Microscopic model.}
To understand microscopically 
the possible mechanism of ferroelectricity in the E-phase we use Monte Carlo (MC) simulations
to study the ground state properties of the following Hamiltonian for manganites
based on the orbitally degenerate 
double-exchange model~\cite{Dagotto01,Efremov04,Hotta03} with one $e_g$-electron per Mn$^{3+}$ ion, 
\begin{eqnarray}\label{ham}
H &=& -\sum_{\mb{i a} \alpha\beta} C_{\mb i, \mb i + \mb a}t^{\mb i \mb a}_{\alpha\beta}
d^\dagger_{\mb i\alpha} d_{\mb i + \mb a \beta}
+ J_\text{AF} \sum_{\mb i \mb a} \mb{S_i}\cdot \mb{S}_{\mb i + \mb a}\\
&& + \lambda\sum_i(Q_{1\mb i}\rho_{\mb i} + Q_{2\mb i} \tau_{x\mb i} + Q_{3\mb i} \tau_{z\mb i})
+ \frac 1 2\sum_{\mb i m} {\kappa_m Q_{m \mb i}^2}, \nn
\end{eqnarray}
where $d^\dagger_{\mb i\alpha}$ is the creation operator for the $e_g$ electron on orbital 
$\alpha = x^2-y^2 (\text{a}), 3z^2-r^2 (\text{b})$; $\mb a = \mb x, \mb y$ is the direction of the 
link connecting the two nearest neighbor Mn sites; and 
$\mb S_{\mb i}$ is the classical unit spin, given by the 
polar angle $\theta_{\mb i}$ and azimuthal angle $\phi_{\mb i}$, representing
the electrons occupying the $t_{2g}$ orbitals on  the $\mb i$-th Mn site. The $t_{2g}$ spins interact directly 
through the antiferromagnetic superexchange $J_\text{AF}$$>$$0$.
$C_{\mb i, \mb j} = \cos \frac{\theta_{\mb i}} 2 \cos \frac{\theta_{\mb j}} 2 + \sin \frac{\theta_{\mb i}} 2 \sin 
\frac{\theta_{\mb j}} 2 e^{-i(\phi_{\mb i}-\phi_{\mb j})}$ is the \emph{double-exchange factor} 
arising due to the large Hund's coupling that projects out the $e_g$ electrons with spin 
antiparallel to $\mb S_i$. $Q_{m \mb i}$ represent the classical adiabatic phonon modes, with stiffnesses 
$\kappa_m$, due to the displacements of the ligand oxygen ions surrounding the $\mb i$-th Mn site. 
The phonon stiffnesses were chosen as follows -- for the Jahn-Teller modes:
$\kappa_1 = 2.0$, $\kappa_2=\kappa_3=1.0$; for the FE mode $\kappa_\text{FE} = 8.0$ and for the rest 
of the modes, $\kappa = 10.0$. Except for the Jahn-Teller modes~\cite{Dagotto01,Hotta03}, this choice 
is driven only by the efficiency of the MC simulations, and it does not reflect the actual frequencies of 
the phonon modes, which are presently unknown. 
We illustrate below how $P$ can be obtained in physical units. 
The third term in Eq.~(\ref{ham}) is the Jahn-Teller coupling with constant $\lambda$ and the $e_g$-orbital operators
$\rho_{\mb i} = d^\dagger_{\mb i a}d_{\mb i a} + d^\dagger_{\mb i b}d_{\mb i b}$, $\tau_{x\mb i} = 
d^\dagger_{\mb i a}d_{\mb i b} + d^\dagger_{\mb i b}d_{\mb i a}$ and 
$\tau_{z\mb i} = d^\dagger_{\mb i a}d_{\mb i a} - d^\dagger_{\mb i b}d_{\mb i b}$. 

\begin{figure}
\begin{minipage}[t]{0.5\columnwidth}
\parbox{40mm}{\includegraphics[clip,width=40mm]{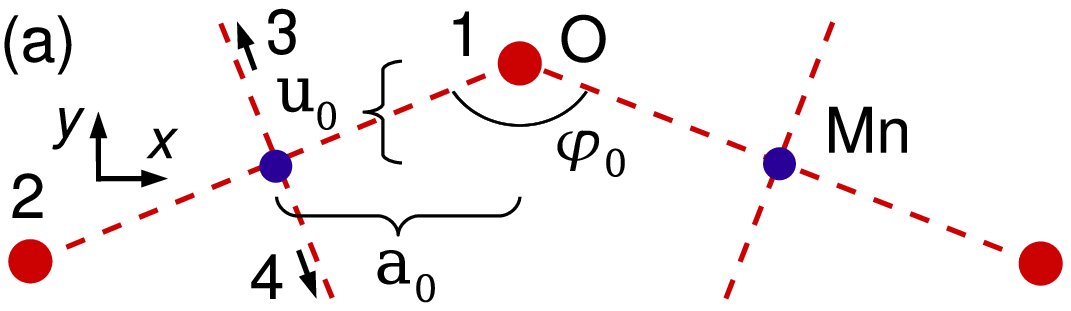}}
\end{minipage}%
\begin{minipage}[t]{0.5\columnwidth}
\parbox{15mm}{\includegraphics[clip,width=15mm]{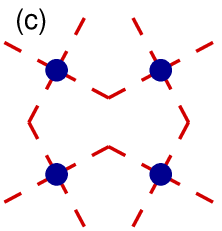}}
\parbox{3mm}{ $\Rightarrow$}
\parbox{15mm}{\includegraphics[clip,width=15mm]{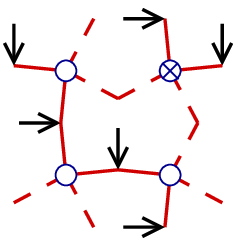}}
\end{minipage}
\vspace{0.2cm}
\begin{minipage}[b]{0.5\columnwidth}
\includegraphics[clip,width=30mm]{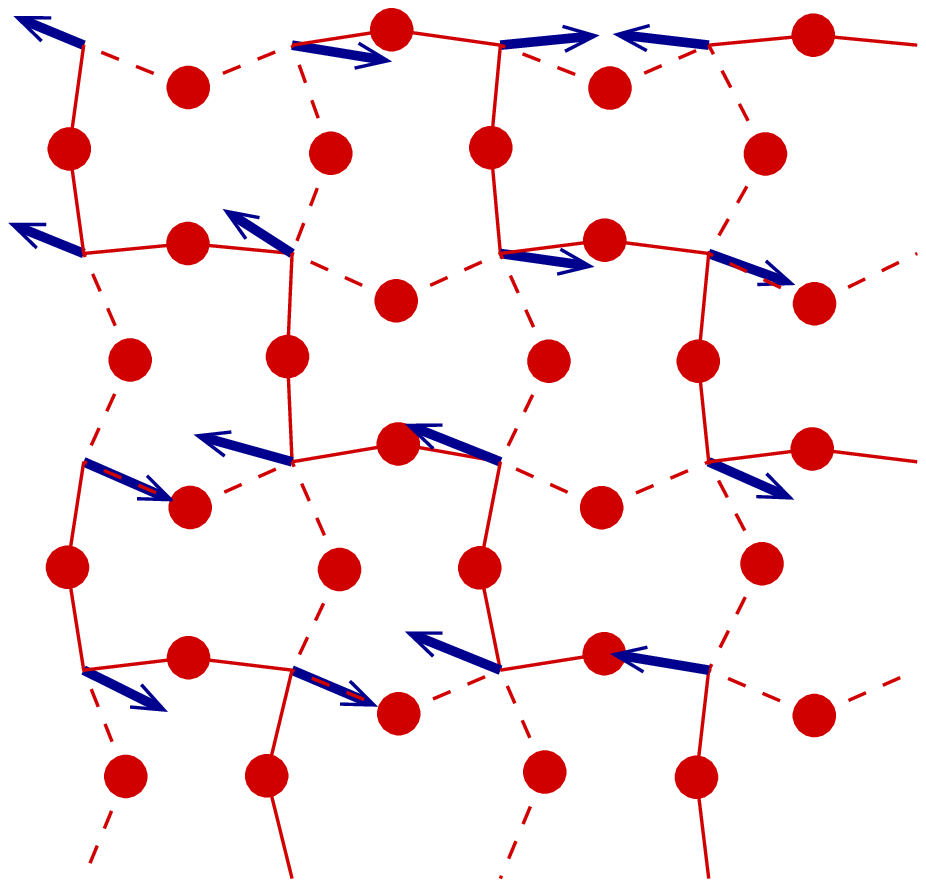}
\hspace{-0.2cm}
\includegraphics[clip,width=8mm]{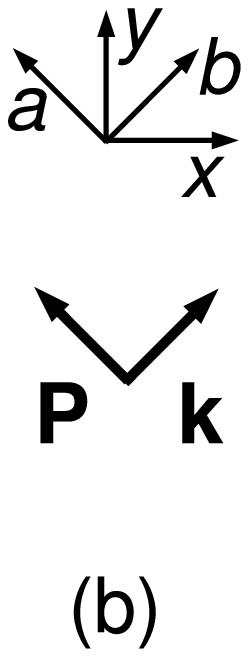}
\end{minipage}%
\begin{minipage}[b]{0.5\columnwidth}
\includegraphics[clip,width=42mm]{P_vs_Phi.eps}
\end{minipage}
\caption{\label{snapshot}(Color online) (a) The starting configuration of a Mn-O-Mn bond. Numbers 1-4 enumerate the O atoms 
surrounding one Mn. (b) A MC snapshot of the IMF E-phase at T=0.01. 
The ferromagnetic zigzag chain links are shown as solid lines. The displacements of the oxygen atoms are 
exaggerated. (c) \emph{Left.} The local arrangement of the Mn-O bonds with disordered Mn spins (full circles).
\emph{Right.} Oxygen displacements (arrows) within the chains of opposite Mn spins (open and crossed circles)
 in the E-phase (see also Refs.\cite{Goto04,Aliouane06}). 
(d) MC results for the polarization at T=0.01 for different values of $J_\text{AF}$.}
\end{figure}

Our improvements over previous approaches to this model are the following.
\emph{First}, we explicitly consider the dependence of the hoping parameters $t^{\mb i\mb a}_{\alpha\beta}$
on the angle of the Mn-O-Mn bond $\varphi_{\mb i\mb a}$. Taking into account only the largest Mn-O $\sigma$-bond 
contribution, we find~\cite{Slater54}
$t^{\mb x}_\text{aa} = t^\mb{y}_\text{aa} = -t\cos^3\varphi$,
$t^{\mb x}_\text{bb} = t^{\mb y}_\text{bb} = -t\cos\varphi /3$,
$t^{\mb x}_\text{ab} = t^{\mb x}_\text{ba} = -t^{\mb y}_\text{ab} = -t^{\mb y}_\text{ba} = -t\cos^2\varphi /\sqrt 3$,
where $t = \frac 3 4 (pd\sigma)^2$ is taken as the energy unit hereafter. 
We neglect the $\varphi$ dependence of $J_\text{AF}$ since it has a smaller energy scale than $t$.
\emph{Second}, $Q_{m \mb i}$'s are defined such that the elastic energy term in~(\ref{ham}) is minimal for 
$\varphi_{\mb i\mb a} \equiv \varphi_0 < 180^\circ$. This allows us to model the initial structural buckling
(GdFeO$_3$-type) distortion present in the orthorhombic perovskites~\cite{Kimura03b}. In particular, the buckling 
mode is defined in two dimensions as
$
Q_{\text{buckle}, \mb i} = y_{\mb i 1} - y_{\mb i 2} - x_{\mb i 3} + x_{\mb i 4} - (-1)^{i_x + i_y} 4 u_0,
$
where $x_{\mb i k}$, $y_{\mb i k}$ are the displacements of the oxygen atoms from their ideal position in
the $180^\circ$ Mn-O-Mn bond, $u_0 = a_0 \cot \frac {\varphi_0} 2$, and $a_0$ is the Mn-O distance in the 
$180^\circ$ Mn-O-Mn bond [see Fig.~\ref{snapshot}(a)]. Since the buckling distortion in the considered perovskites 
is present at temperatures well above the magnetic transitions, we consider $\varphi_0$ as a fixed 
parameter of the model. For \ho, $\varphi_0$ is close to 144$^\circ$~\cite{Kimura03b}. 
As shown below, this distortion plays a crucial role in generating ferroelectricity in the E-phase.
Introducing $\varphi_0< 180^\circ$ effectively reduces the symmetry of the Hamiltonian~(\ref{ham}),
although it is still invariant with respect to the inversion symmetry centers located at the every Mn site. 
The FE polarization emerges only due to the spontaneous symmetry breaking caused by
the E-phase magnetic order. 

\emph{Monte Carlo results.} 
Many results of the model~(\ref{ham}) regarding magnetic and orbital order, without the modifications mentioned 
above, are reported in Ref.~\cite{Hotta03}. Our present treatment essentially confirms those results. The ground 
state phase diagram in two dimensions contains the ferromagnetic phase, E-phase, and the Ne\'el-like G-phase. 
Here we focus our attention on the new results directly related to the ferroelectricity of the E-phase. 
A typical low-temperature E-phase MC snapshot is shown in Fig.~\ref{snapshot}(b). 
In accordance with the absence of spin-orbit interaction in Hamiltonian~(\ref{ham}), it is invariant with respect 
to collective spin rotations, therefore
the preferred spin direction is chosen randomly in our MC simulations. However, the ferromagnetic zigzag 
chains are clearly established in the ground state. As is seen from the snapshot, the double-exchange physics
plays a crucial role in the formation of the FE state. Due to the factor $C_{\mb i, \mb i + \mb a}$ the electron 
hopping is prohibited between Mn atoms with opposite $t_{2g}$ spins. Hence, the displacement of the corresponding 
oxygen atom perpendicular to the Mn-Mn bond (these displacements are not Jahn-Teller active) is only due to the 
elastic energy, which favors the bond angle $\varphi_0$. 
On the contrary, hopping is allowed along the ferromagnetic zigzag chains~\cite{insulator}. 
Since hopping energy is minimal for the $180^\circ$ bond, the competition 
between the hopping and elastic energies generally results in a bond angle $\varphi$, such that
$\varphi_0 < \varphi < 180^\circ$ [see Fig.~\ref{snapshot}(c)]. 
Since $\varphi$ only depends on the nature of the bond (ferromagnetic \emph{vs}.
antiferromagnetic), the direction of the oxygen displacements is the same in all zigzag chains, even though 
neighboring chains have opposite spin. This leads to the overall coherent displacement 
of the center of mass of 
the O atoms with respect to the Mn sublattice, similarly as proposed in Ref.~\cite{Aliouane06} 
for the field-induced phase of TbMnO$_3$. 
It is easy to see from Fig.~\ref{snapshot}(b, c) that the resulting FE polarization points
along the diagonal connecting the next-nearest-neighboring Mn atoms, \emph{i.~e.} the $a$-axis in the orthorhombic 
setting, in perfect agreement with Landau theory. 

\emph{The value of $P$}.
Figure~\ref{snapshot}(d) shows the calculated absolute value of the polarization $P_\text{calc}$, defined 
as the oxygen displacement per 
one ``unit cell'' containing one Mn and two O atoms averaged over MC steps, plotted versus $\varphi_0$ for
different values of $J_{\text{AF}}$. 
$P$ vanishes for $\varphi_0 = 180^\circ$ for all values of $J_\text{AF}$. In this case both the hopping and elastic 
energies are optimal when $\varphi = \varphi_0 = 180^\circ$, the same for ferromagnetic and antiferromagnetic bonds. 
To demonstrate that this is also consistent with the symmetry arguments of Landau theory, we note that the 
symmetry of the lattice
with $180^\circ$ bonds is higher than that with $\varphi_0 < 180^\circ$. Particularly, additional inversion centers are 
located at the O sites if $\varphi_0 = 180^\circ$. While the E-type magnetic structure breaks inversion 
symmetry for centers located at the Mn sites, it is invariant with respect to the O-site centers, and thus the magnetic
phase transition cannot induce FE order. In terms of the Landau potential~(\ref{pot}), the coefficients $c$ and 
$d$ must be zero in this case. 

The MC calculated value of the polarization $P_\text{calc}$ for  $\varphi_0 \approx 145^\circ$ corresponding to \ho\ 
reaches 0.08. It can be shown from the model~(\ref{ham}), that the FE displacements scale as $t/\kappa_\text{FE}a_0$.
In the simulations we set $t=1$ and $a_0=1$, and the polarization in the physical units, in the point charge 
approximation, is given by
$
P = \frac{12\, e\, t\, \kappa_\text{FE}}{\kappa'_{\text{FE}}\, a_0\, V_0} P_\text{calc}, 
$
where $e$ is the elementary charge, $\kappa'_{\text{FE}}$ is the FE phonon stiffnesses in the physical units, and $V_0$ is 
the unit cell volume. Using $a_0 = 2.35$~\AA, $V_0=226$~\AA$^3$~\cite{Munoz01}, 
$t = 0.1 - 0.5$~eV~\cite{Dagotto01}, 
$\kappa'_\text{FE} = 1 - 5$~eV/\AA$^2$~\cite{Sergienko06}, we obtain a range of values for $P$ between
0.5 and 12 $\mu$C/cm$^2$, substantially larger than the $P$ observed in helical
IMF's. We can also estimate $P$ from the known experimental results for TbMnO$_3$. Its crystal
structure is modulated in the collinear sine-wave phase with wave vector $k_L = 2 k_M$ where $k_M=0.29$ is the 
magnetic wave vector. The displacement of the O atom $n$ in a Mn-O chain running along the $x$- or $y$-direction
[see Fig.~\ref{snapshot}(b)]
is given by $\delta \mb r_n =\delta \mb r_0 (-1)^n \{\cos \pi k_M+\cos[(2n+1)\pi k_M+2\alpha]\}$, for the magnetic 
structure defined by $S_a=S_c=0$, $S_b = S_b^0\cos(n\, k_M + \alpha)$~\cite{Sergienko06}. 
If $k_M \ne 1/2$, then $\sum_n \delta r^a_n = 0$ and the state is paraelectric. The E-phase corresponds to 
$k_M = 1/2$, $\alpha = 3\pi/4$, and hence $\delta \mb r_n \equiv \delta \mb r_0$ represents a 
coherent displacement within one chain. While the $b$ and $c$ components of $\delta \mb r_0$ have different
signs in different chains, $\delta r^a_0$'s are the same. Considering the TbMnO$_3$ value 
$\delta r_0 = 10^{-2}$~\AA~\cite{KimuraPrivate} as an estimate for $\delta r^a_0$ in \ho, we obtain 
$P \approx 2\, \mu$C/cm$^2$ consistent with the calculated values.

\emph{Summary}. We argue that the symmetry of the spin zigzag chain magnetic E-phase in orthorhombic perovskites with 
buckling distortion of the oxygen octahedra allows the formation of a polar axis along the $a$-axis. The 
microscopic mechanism of ferroelectricity is independent of spin-orbit coupling, and $P$ 
can potentially be up to two orders of magnitude higher than that in the helical IMF's. Note that the 
ideas described here are general, and the E-phase provides just one example. In fact, the mechanism is similar to the 
recently proposed explanation for the ferroelectricity found in 
TbMnO$_3$ at high fields~\cite{Aliouane06}, and our effort provides a firm theoretical basis for that interesting scenario.
As to the E-phase nickelates, the microscopic model considered here is not applied to them directly~\cite{Hotta04}. 
However, we expect that the same interplay between the optimization of the electron hopping and the elastic
energy in the presence of the buckling distortion will lead to ferroelectricity. 

We thank M. Angst, D. Argyriou, T. Egami, D. Mandrus, and D. Singh for inspiring discussions, as well as 
T. Kimura and B. Lorenz for providing us with their unpublished results.
Research at ORNL is sponsored by the Division of Materials Sciences and Engineering,
Office of Basic Energy Sciences, U.S. Department of Energy, under contract DE-AC05-00OR22725 with ORNL, 
managed and operated by UT-Battelle, LLC. 
This work is also supported in part by NSF DMR-0443144 and NSF DMR-0072998.
We have used the SPF software developed at ORNL (http://scicompforge.org/spf).

\emph{Note added.} Following our prediction, B. Lorenz \emph{et al.}~\cite{Lorenz06} reported on experimental evidence
of ferroelectricity in perovskite \ho\ and YMnO$_3$. For the latter, the E-phase was proposed in Ref.~\cite{Zhou06}.


\begin{thebibliography}{00}
\bibitem{MostNote} 
Although the E-phase is a special case of the simple sine-wave structure, the continuum
theory considered in Ref.~\cite{Mostovoy06}, which predicts $\mb P = 0$, is not applicable to modulation vectors
at the edge of the Brillouin zone.
\bibitem{insulator} Although virtual hopping is allowed along the zigzag chains, the E-phase is a band 
insulator~\cite{Hotta03}.
\end{thebibliography}
\end{document}